\documentclass[
aps,%
12pt,%
final,%
notitlepage,%
oneside,%
onecolumn,%
nobibnotes,%
nofootinbib,%
superscriptaddress,%
noshowpacs]%
{revtex4}%
\usepackage{graphics}
\usepackage[english]{babel}
\usepackage{amssymb,amsmath}

\begin{document}

\title{Associative production of $B_c$ and $\bar D$ mesons at LHC.}
\author{\firstname{A.~V.}~\surname{Berezhnoy}}
\email{Alexander.Berezhnoy@cern.ch}
\affiliation{SINP of Moscow State University, Russia}

\author{\firstname{A.~K.}~\surname{Likhoded}}
\email{Anatolii.Likhoded@ihep.ru}
\affiliation{Institute for High Energy Physics, Protvino, Russia}

\author{\firstname{A.~A.}~\surname{Martynov}}
\email{aler1x@yandex.ru}
\affiliation{Physical Department of Moscow State University, Russia}

\begin{abstract}
It is shown that the study of correlations
  in the associative production of $B_c$ and $\bar D$ mesons at LHC  
allows one to obtain essential  information about the $B_c$ production mechanism.

\end{abstract}

\maketitle

\section{Introduction}

Recent measurements of $B_{c}$ meson mass and lifetime in CDF \cite{Bc_CDF}
and D0 \cite{Bc_D0} experiments  are the first steps in the experimental research of quarkonia with open flavor. The measurement results  are in a good agreement with the theoretical predictions for the $B_c$ 
mass~\cite{Bc_spectroscopy_Kiselev, Bc_spectroscopy_Eichten, Bc_spectroscopy_Galkin}:
$$m_{B_c}^{\rm CDF}=6.2756\pm 0.0029(\textrm{stat.})
\pm 0.0025(\textrm{sys.}) \;\textrm{GeV}, $$
$$m_{B_c}^{\rm D0}=6.3000\pm 0.0014(\textrm{stat.})
\pm 0.0005(\textrm{sys.}) \;\textrm{GeV},
$$
$$ \qquad  m_{B_c}^{\rm theor}=6.25\pm 0.03 \;\textrm{GeV};$$
as well as for the decay time~\cite{Bc_decay_SR_Kiselev,Bc_decay_PM_Kiselev}:
$$\tau_{B_c}^{\rm CDF}=0.448^{+0.038}_{-0.036} (\rm stat.)\pm 0.032 (\rm sys.)\;\textrm{ps},$$
$$\tau_{B_c}^{\rm D0}=0.475^{+0.053}_{-0.049} (\rm stat.)\pm 0.018 (\rm sys.)\;\textrm{ps},$$
$$\tau_{B_c}^{\rm theor}=0.48 \pm 0.05 \;\textrm{ps}.$$

Unfortunately, the experimental estimations of the cross section value were not published. 
Thus, the  mechanism of $B_c$ meson production can not be understood from the obtained data due to poor experimental statistics, as well as due to large uncertainties in the theoretical predictions.
Only the planed experimental research at LHC, where about $10^{10}$ events with $B_c$ mesons per year are expected, will improve the situation. This huge amount of events will allow to obtain the information on the production cross section distributions, on the decay branching fractions, and in some cases, on the distributions of decay products.  Obviously, the new experimental data on  $B_c$ meson from LHC experiments have to stimulate new theoretical studies.

At first sight  the production of $B_c$ meson can be represented as the heavy $\bar b$ quark production followed by the fragmentation of $\bar b$ quark into $B_c$ meson by analogy with the production of $B$ mesons, at least for transverse momenta larger than the $B_c$ meson mass\footnote{Within fragmentation approach it is assumed that $b \bar b$ phase space can be factorized from the total phase space. This assumption is not valid for final state $B_c+b+\bar c$ at small momenta, because $c$ quark   is massive.}. 
But, as it  was  shown in the framework of pQCD, there  is a large nonfragmentation contribution into this process~\cite{Bc_hadronic_production, Bc_production_Chang_Chen_Han_Jiang, Bc_production_Kolodziej_Leike_Ruckl,Bc_production_Baranov}.   The nonfragmentation terms in the production amplitude enlarge the total cross section and increase the ratio between yields of $B_c^*$ and $B_c$ states by a factor of two even at large transverse momenta.

However, our previous analysis showed that it is quite difficult to obtain a clear experimental evidence of the nonfragmentation contribution into the production~\cite{Bc_hadronic_production}.
In this article we try to fill the gap in our understanding of $B_c$ production  and show that the study of correlations in  the associative production of $B_c$ and $\bar D$ meson at LHC  
could be an essential information source of $B_c$ production mechanism. The cross section distributions of the $B_c$ meson production  in hadronic and gluonic interactions have been obtained by us within the numerical calculation method developed in our previous studies~\cite{Bc_hadronic_production}.

\section{Fragmentation and recombination contributions into $B_c$ production}

The $B_c$ production amplitude within the discussed approach can be subdivided into two parts: the hard production of two heavy quark pairs calculated  in the 
framework of perturbative QCD and the  soft nonperturbative 
binding of $\bar b$ and $c$ quarks into quarkonium  described by nonrelativistic  wave function. The production amplitude within this approach can written as  follows:
 \begin{equation}
A^{SJj_z}=\int T^{Ss_z}_{b\bar b c \bar c}(p_i,k(\vec q))\cdot
\left (\Psi^{Ll_z}_{\bar b c}(\vec q) \right )^* \cdot
C^{Jj_z}_{s_zl_z} \frac{d^3 \vec q}{{(2\pi)}^3},
\end{equation}
where $T^{Ss_z}_{b\bar b c \bar c}$ is an amplitude of the hard production of two heavy quark pairs;\\
$\Psi^{Ll_z}_{\bar b c}$ is the quarkonium wave function;\\
$J$ and $j_z$ are the total angular momentum and its projection on $z$-axis in the $B_c$ rest frame;  \\
$L$ and $l_z$ are the orbital angular momentum of $B_c$ meson and its projection on $z$ axis;\\
$S$ and $s_z$ are $B_c$ spin and its projection;\\
$C^{Jj_z}_{s_zl_z}$ are Clebsh-Gordon coefficients;\\
$p_i$ are four momenta of  $B_c$ meson, $b$ quark and  $\bar c$ quark;\\
$\vec q$  is three momentum of $\bar b$ quark in the  $B_c$ rest frame
(in this frame $(0,\vec q ) = k(\vec q)$).

Under the assumption of small dependence of $T^{Ss_z}_{b\bar b c \bar c}$ on $k(\vec q)$ 
\begin{equation}
A \sim 
\int d^3q\,\Psi^*({\vec q})\left\{ \bigl.T(p_i,{\vec q}) \bigr|_{\vec q=0}+
\bigl.{\vec q}\frac{\partial}{\partial {\vec q}} T(p_i,\vec q) \bigr|_{\vec q =0}  +  \dotsb \right\}
\end{equation}
and particularly for the $S$ wave states
\begin{equation}
A \sim R_S(0) \cdot \bigl. T_{b\bar b c \bar c}(p_i) \bigr|_{\vec q=0}, 
\end{equation}
where $R_S(0)$ is a value of radial wave function at  origin.
\begin{figure}[!t]
\centering
\resizebox*{1.05\textwidth}{!}{
\includegraphics{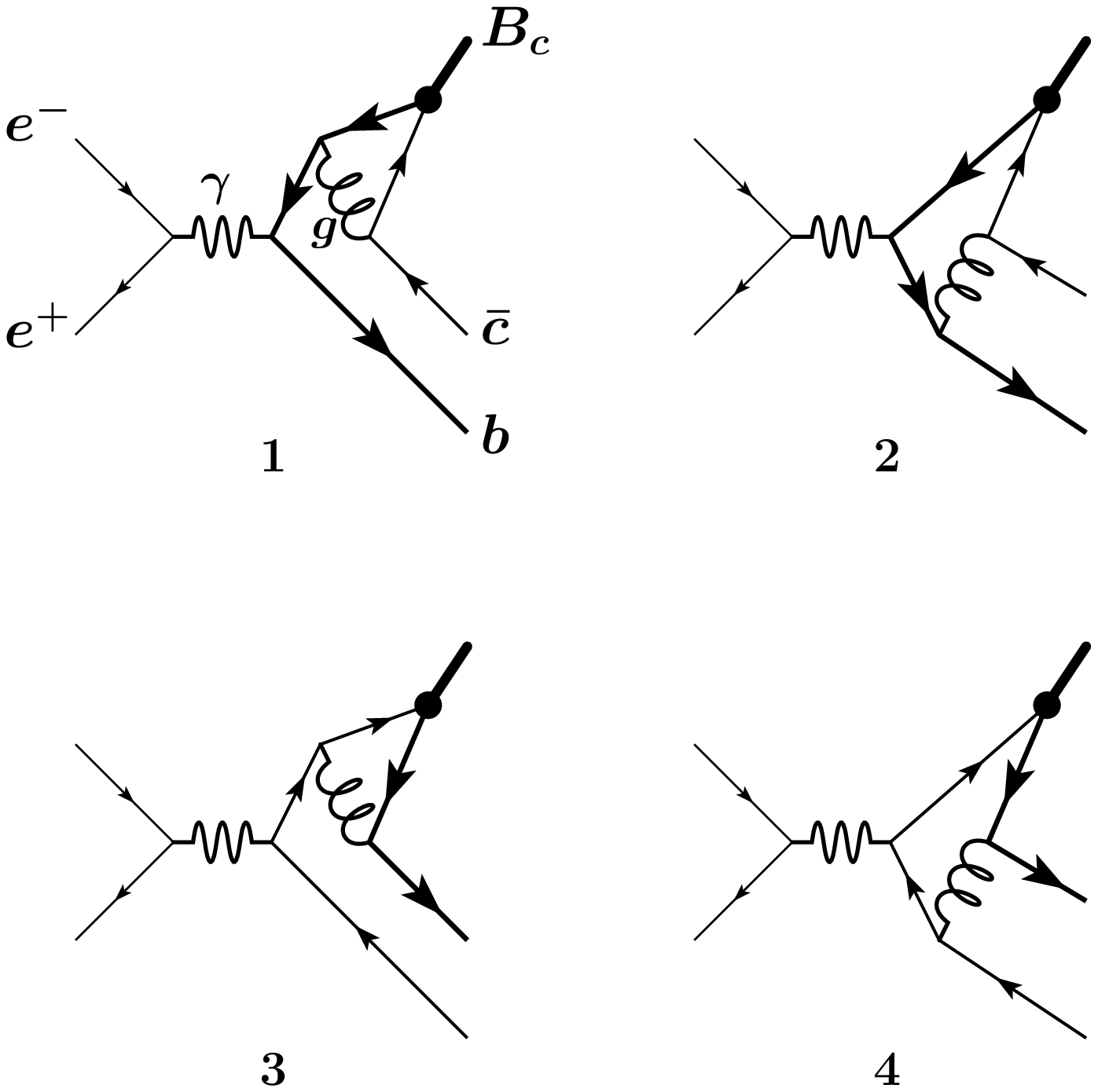}
\hfill
\includegraphics{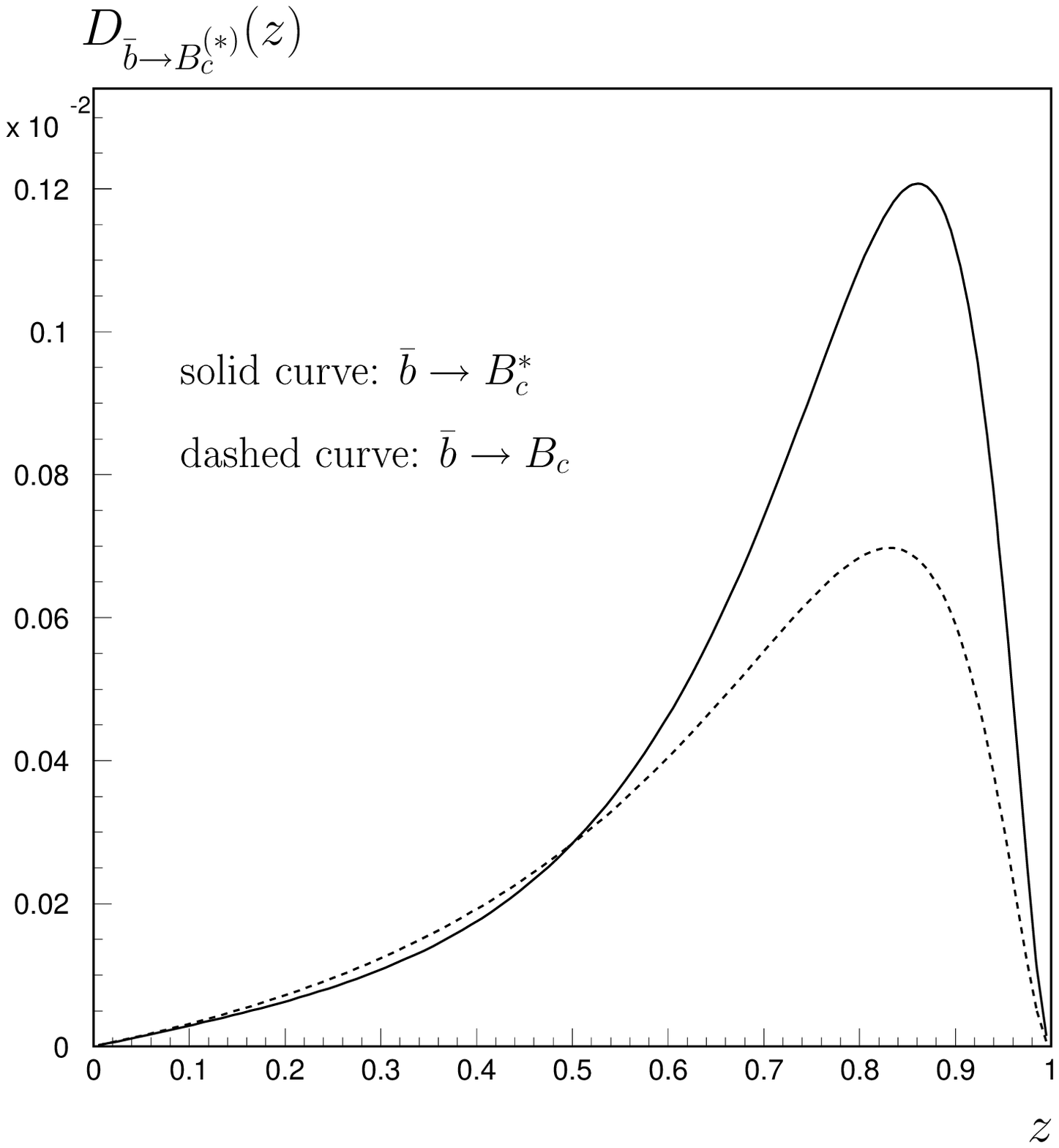}
}
\\
\parbox[t]{0.47\textwidth}{
\caption{ The leading order diagrams for the process $e^+e^-\to B_c + b +\bar c$.
\hfill}
\label{fig:diagr_ee}}
\hfill
\parbox[t]{0.47\textwidth}{
\caption{The fragmentation functions for the processes $\bar b\to B_c^*$ (solid curve) and $\bar b \to B_c$ (dashed curve) calculated within pQCD.}\label{fig:fragfunc}}
\end{figure}

The  calculations within the discussed technique are the most simple for  the process of $B_c$  production in the  $e^+e^-$ annihilation,  where only four leading diagrams contribute to the process amplitude (see Fig.~\ref{fig:diagr_ee}).  As it was shown in~\cite{Bc_fragmentation_S_wave_Braaten, Bc_fragmentation_S_wave_Kiselev, Bc_production_Kolodziej_Leike_Ruckl}, the diagrams (3) and (4) are suppressed as $m_c^2/m_b^2$ for the $S$ wave state production due to the large gluonic virtuality. The special choice of gluonic field gauge allows to neglect the contribution of diagram (2)  at large energies $\sqrt{s} \gg M_{B_c}$. The remained contribution  of diagram (1) can be naturally interpreted as the $\bar b$ quark production  followed by the fragmentation of $\bar b$ quark into $B_c$ meson.
Thus,  in $e^+e^-$ annihilation at large energies 
the consideration of leading diagrams for the $B_c$ meson production leads the well known 
factorized formulas  for the  cross section distribution over $z=2 E_{B_c}/\sqrt{s}$ and over $B_c$ transverse momentum $p_T$: 
\begin{equation}
\frac{d\sigma}{dz} = \sigma_{b \bar b}\cdot D(z),
\end{equation}
\begin{equation}
\frac{d\sigma_{B_c}}{dp_T}=\int\limits^1_{2p_T/\sqrt{s}}
\frac{d\sigma_{b \bar b}}{dk_T}
\left(\frac{p_T}{z}\right)\frac{D(z)}{z} dz,
\label{eq:frag_pT}
\end{equation}
where $k_T$ is a transverse momentum of $\bar b$-quark.

  The analytical forms of fragmentation functions for $S$ wave states are known from 
\cite{Bc_fragmentation_S_wave_Braaten, Bc_fragmentation_S_wave_Kiselev}
 (see Fig.~\ref{fig:fragfunc}):
\begin{multline}
       D_{\bar b \to B_c}(z)=
\frac{2\alpha_s^2 |R_S(0)|^2}{81\pi m_c^3}
\frac{rz(1-z)^2}{(1-(1-r)z)^6}
(6-18(1-2r)z+(21-74r+68r^2)z^2
\\
-2(1-r)(6-19r+18r^2)z^3+
3(1-r)^2(1-2r+2r^2)z^4),
\label{eq:frag_ps}
\end{multline}
\begin{multline}
D_{\bar b \to B_c^*}(z)=
\frac{2\alpha_s^2 |R_S(0)|^2}{27\pi m_c^3}
\frac{rz(1-z)^2}{(1-(1-r)z)^6}
(2-2(3-2r)z+3(3-2r+4r^2)z^2\\
-2(1-r)(4-r+2r^2)z^3+
(1-r)^2(3-2r+2r^2)z^4),
\label{eq:frag_vct}
\end{multline}
where  $\alpha_s$  is a strong coupling constant and $r=\frac{m_c}{m_b+m_c}$.

\begin{figure}[!t]
{\centering \resizebox{0.7\textwidth}{!}
{\includegraphics{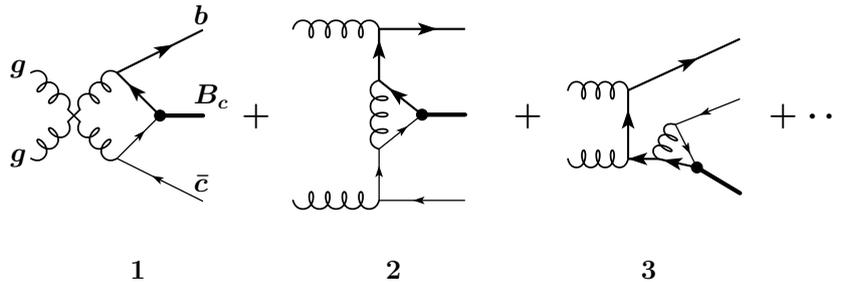}}
}
\caption{The leading order diagrams for the process $gg\to B_c + b +\bar c$.
\hfill \label{fig:diagr_gg}}
\end{figure}

\begin{figure}[!t]
\centering
\resizebox*{1.05\textwidth}{!}{
\includegraphics{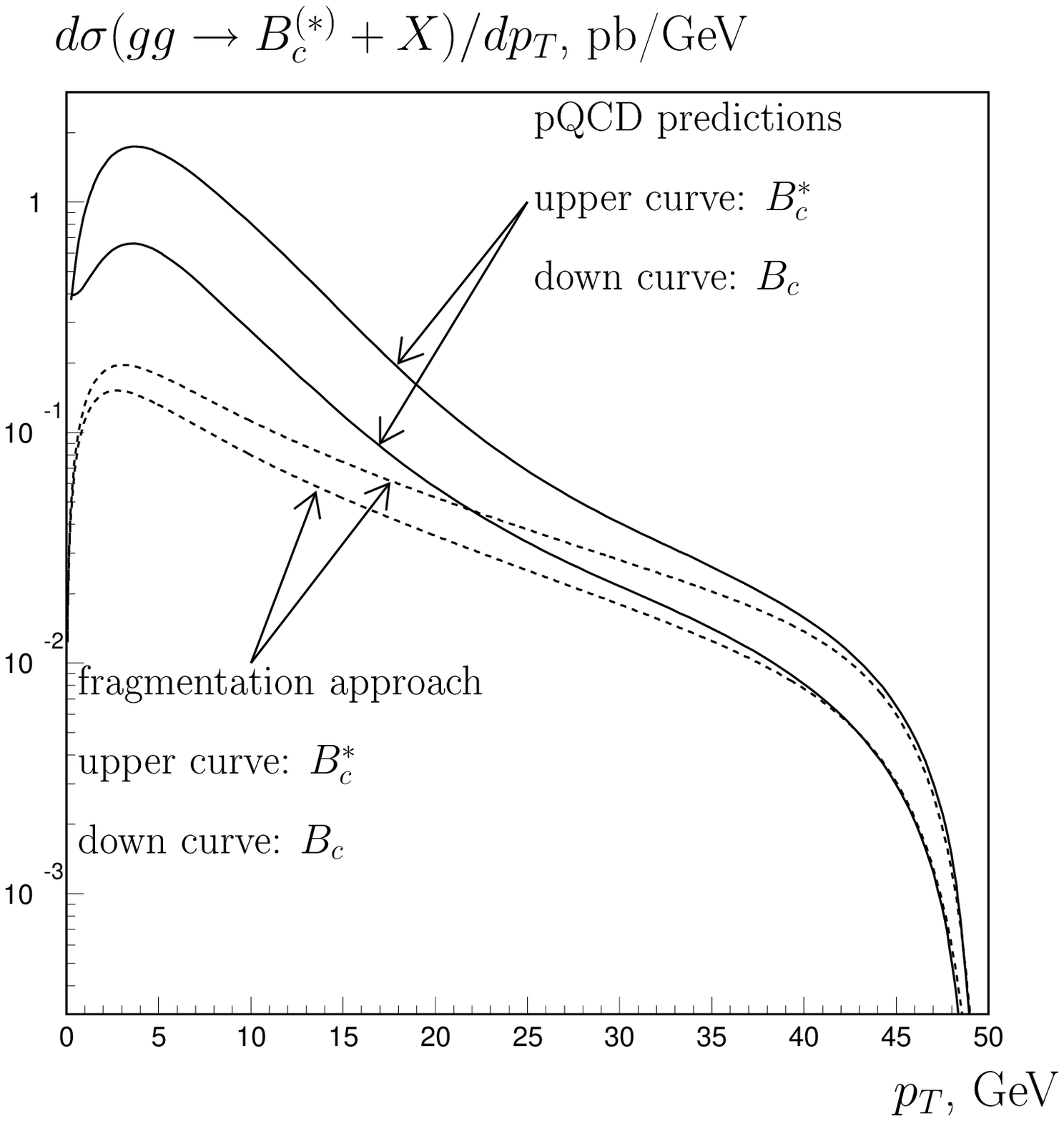}
\hfill
\includegraphics{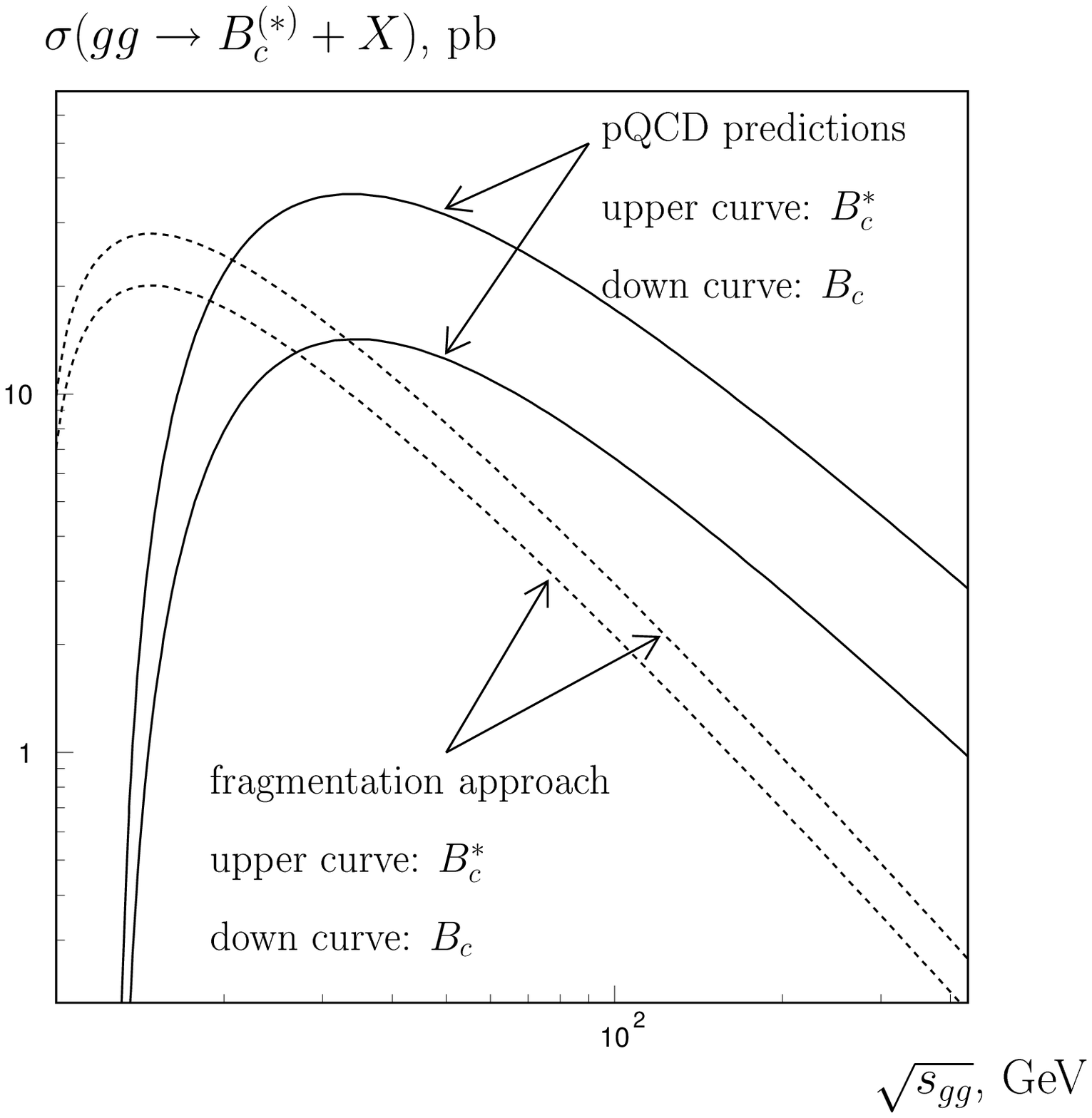}
}
\\
\parbox[t]{0.47\textwidth}{
\caption{ The cross section distribution on  transverse momenta for the $B_c$ gluonic production at interaction energy 
$\sqrt{s_{gg}}=100$~GeV   obtained within pQCD~\cite{Bc_hadronic_production} in comparison with the fragmentation model prediction (see formula~(\ref{eq:frag_pT}) for the fragmentation model).   \hfill}
\label{fig:pT_gg}}
\hfill
\parbox[t]{0.47\textwidth}{
\caption{The cross section value dependence on the interaction energy for the  gluonic $B_c$ production~\cite{Bc_hadronic_production}.}\label{fig:s_hat_gg}}
\end{figure}

\begin{figure}[!t]
\centering
\resizebox*{1.\textwidth}{!}{
\includegraphics{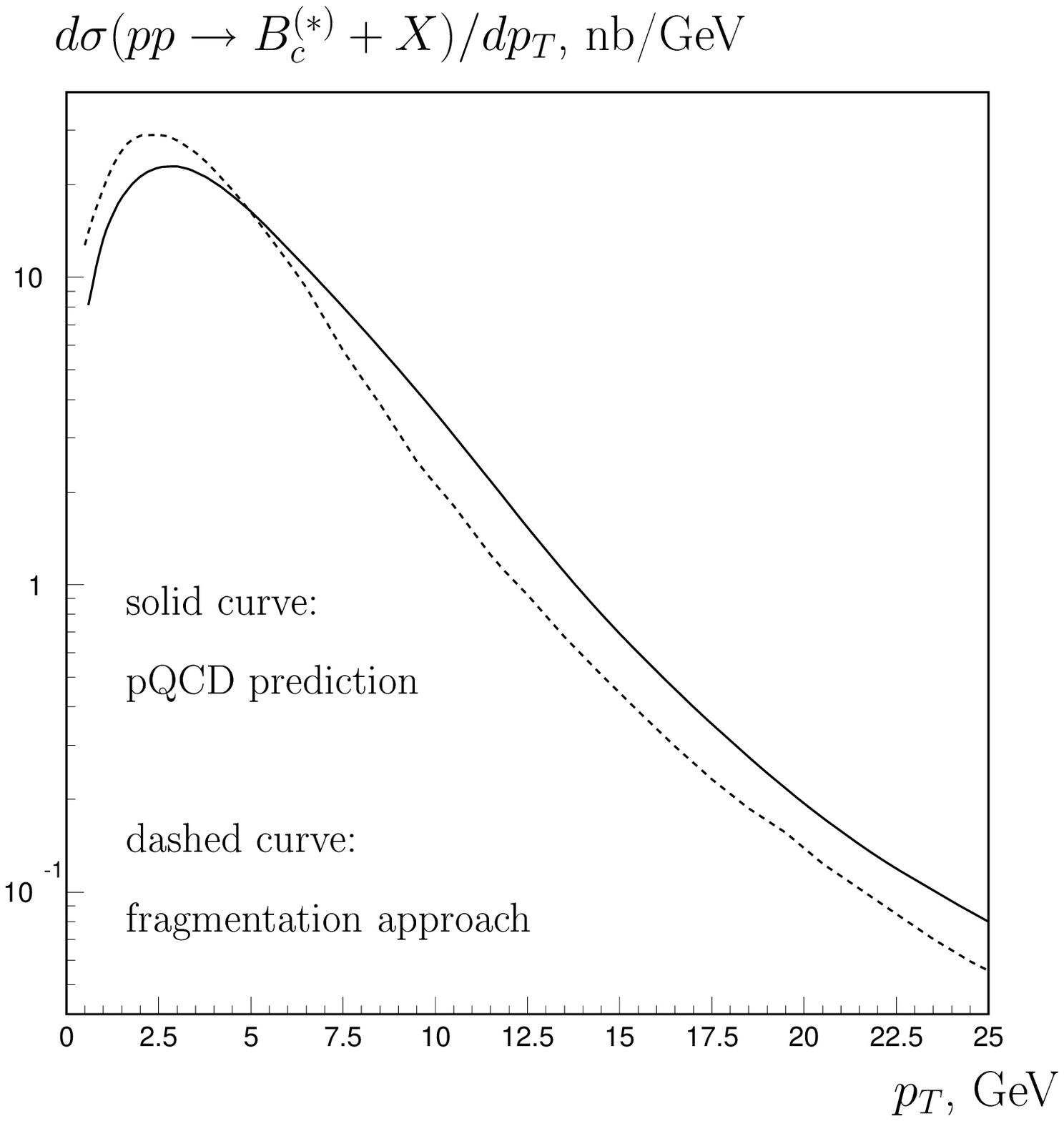}
\hfill
\includegraphics{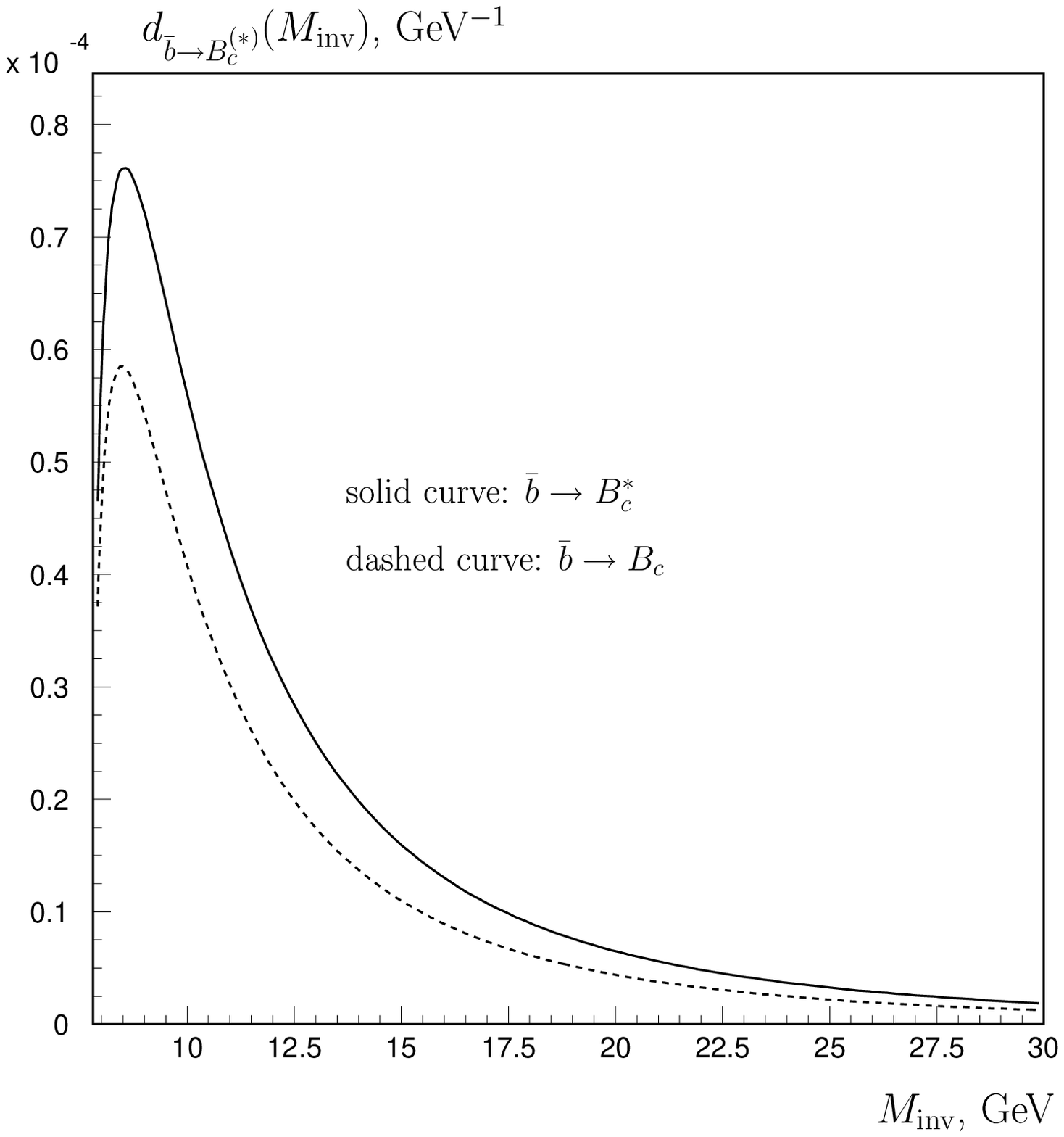}
}
\\
\parbox[t]{0.47\textwidth}{
\caption{The cross section distribution over the $B_c$ transverse
momentum for the process $pp \to B_c +X$ at $\sqrt{s}=14$~TeV within partonic approach~(\ref{eq:convolution}) (see also~
\cite{Bc_hadronic_production}).\hfill}
\label{fig:pT_pp}}
\hfill
\parbox[t]{0.47\textwidth}{
\caption{The normalized cross section distributions over the invariant mass of $B_c$ ($B_c^*$) and $c$-quark within fragmentation approach.}\label{fig:frag_minv}}
\end{figure}

The graphics  presented in figures~\ref{fig:fragfunc}-\ref{fig:BcDcostheta_pp} have been obtained for the following parameter values:
$ \alpha_s=0.2$, $ m_c=1.5 \; \textrm{GeV}$,
$ m_b=4.8 \; \textrm{GeV}$,
$ R_S(0)=2.14 \; \textrm{GeV}^3$.  As it is shown in~\cite{Bc_production_Chang_Wu_uncert} the uncertainties in these parameters leads to the quite large uncertainties in the absolute cross section values. In this work we study the production characteristics which weakly depend on these parameters: the cross section distribution shapes and the ratio between $B_c^*$ and $B_c$ yields.

 The relative yield of $B_c^*$ and $B_c$ in $e^+e^-$ annihilation obtained within pQCD calculation  $R^{B_c}_{e^+e^-}=\sigma(B_c^*)/\sigma(B_c)\sim 1.4$. Thus the naive spin counting which fairly predicts this ratio for $B^*$  and $B$ ($R^{B}_{e^+e^-}\sim 3$) can not be applied to $B_c$ and $B_c$ 
production.~\footnote{Within the valence quark approximation  
$|B^*\rangle = \left\{ b_\downarrow \bar q_\downarrow, \; 
(b_\uparrow \bar q_\downarrow + b_\downarrow\bar q_\uparrow)/\sqrt{2}, \; 
b_\uparrow \bar q_\uparrow \right\}$ and
$|B\rangle = (b_\uparrow \bar q_\downarrow - b_\downarrow\bar q_\uparrow)/\sqrt{2}$. Thus, if it is supposed that all possible spin combinations are produced with the same probability, then $\sigma(B^*)/\sigma(B)=3$.
}

At first sight  it would be reasonable to assume that for the gluonic $B_c$ production the fragmentation mechanism is also dominant at least from  transverse momenta larger than $B_c$ mass. But as it was shown in \cite{Bc_hadronic_production, Bc_production_Chang_Chen_Han_Jiang, Bc_production_Kolodziej_Leike_Ruckl,Bc_production_Baranov} the other mechanism essentially contribute to this process practically all over the phase space. The thing is that in the leading order of pQCD the gluonic production described by 36 diagrams (see Fig.~\ref{fig:diagr_gg}) and only a small part of them belongs to the  fragmentation type. The other diagrams  can  not be interpreted as $\bar b$ quark production  followed by the fragmentation into $B_c$ meson (see  for example the diagram (1) in Fig.~\ref{fig:diagr_gg} ).
One can say that these (recombination) diagrams correspond to  the "independent" creation of $c\bar c$ and $b \bar b$ pairs. In spite of the suppression by a factor  $1/p_T^2$ comparing with the fragmentation diagrams,   their contribution into amplitude can not be neglected even at relatively large transverse momenta. As one can see in Fig.~\ref{fig:pT_gg} the fragmentation approach is valid only at transverse momenta larger than $5\div 6$ masses of $B_c$.  This behavior does not contradict the factorization theorem~\cite{factorization_theorem}, which predicts the fragmentation dominance at limit of infinite transverse momenta, but says nothing about 
transverse momentum value at which the nonfragmentation contribution becomes negligible.

The  total gluonic cross section predicted  using full set of leading order diagrams essentially differ from the fragmentation approach in  absolute value as well as in shapes of  interaction energy dependence (see Fig.~\ref{fig:s_hat_gg}). It is worth to note that near the threshold the fragmentation prediction values are larger than the prediction values obtained within pQCD calculation because of  the incorrect phase space counting in fragmentation approach.

To obtain the cross section estimations for real hadronic experiments we need to convolute the partonic cross sections with the parton density functions:
\begin{equation}
\sigma_{\rm hadronic}(s)=\int \int \sigma_{ij} (\hat s_{ij},Q_R)  f_i(x_i,Q_F) f_j(x_j,Q_F) 
d x_i d x_j 
\label{eq:convolution}
\end{equation}

As it is predicted within pQCD~\cite{Bc_hadronic_production,Bc_production_Chang_Chen_Han_Jiang}, about 90 \% of the $B_c$ mesons at LHC energies will be produced in the gluonic fusion (Fig.~\ref{fig:diagr_gg}). Therefore in our pQCD calculations we can neglect the other partonic subprocess.

The convolution of the gluonic subprocess cross section with the gluonic structure functions partially hides the  differences between the pQCD predictions and the fragmentation approach.  For example,  one can see in Fig.~\ref{fig:pT_pp} that at hadronic level the difference in shapes of $p_T$ distributions  is not so obvious, as at partonic level.~\footnote{To estimate the $B_c$ hadronic production cross section the CTEQ6L parametrization of gluonic structure function has been used on the scale $Q_F=10$~GeV~\cite{CTEQ}.}
The predicted difference of the hadronic cross section values is about of 2. Obviously, such a difference  is not  essential for the  calculations  of forth order on $\alpha_s$.
Nevertheless, the relative yield of $B_c^*$ and $B_c$ does not depend on  $\alpha_s$ and could indicate the production mechanism. Even in the kinematical region where the fragmentation model could be applied   the value of $R_{\rm hadr}=\sigma_{\rm hadr}(B_c^*)/\sigma_{\rm hadr}(B_c)$ predicted within pQCD is about 2.6 instead of 1.4 obtained within the fragmentation approach. To measure this value one need to detect $B_c^*$ meson which with unit probability decays into $B_c$ and photon.   However, it is quite difficult to detect such a process experimentally due to the expected small mass
difference  between $B_c^*$ and $B_c$ mesons~\cite{Bc_spectroscopy_Kiselev}: 
\begin{equation}
\Delta M =M_{B_c^*}-M_{B_c}= 65 \pm 15 \; \textrm{MeV}.
\end{equation}
 Therefore in laboratory system  the maximum energy of emitted photon is
$$\omega_{max}=\left(\gamma+\sqrt{(\gamma^2-1)}\right)\Delta M,$$
where $\gamma$ is $B_c^*$ $\gamma$ factor. 
Even for $B_c^*$ with energy  $\sim 30$ GeV $\omega_{max}\sim 0.7$ GeV. Thus one can conclude that there is no certainty  that the method based on separation of $B_c^*$ from $B_c$ can be used to study the $B_c$ production mechanism. 
This is why in the next chapter we suggest another way to research the $B_c$ production mechanism.

\section{The cross section distribution on the invariant mass of $D$ and $B_c$ mesons}

Within fragmentation approach the shape of the cross section distribution over the invariant mass of $B_c$ and $\bar c$-quark is roughly determined by $\bar b$-quark virtuality (see  diagram (1) in Fig.~\ref{fig:diagr_ee}). 
This is why the distribution should be relatively narrow. Our analytical calculations confirm this supposition.  Let us denote by $d_{\bar b\to B_c}(M_{B_c+\bar c})$ and $d_{\bar b\to B_c^*}(M_{B_c+\bar c})$ the 
normalized cross section distributions over the invariant mass of $B_c$ ($B_c^*$) and $c$-quark within fragmentation approach:
\begin{equation}
d_{b\to B_c}(M_{B_c+\bar c})=\frac{1}{\sigma_{b \bar b}}
\frac{d\sigma^{\rm frag}_{B_c}}{d M_{B_c+\bar c}}, 
\end{equation}
\begin{equation}
d_{\bar b\to B_c^*}(M_{B_c+\bar c})=\frac{1}{\sigma_{b \bar b}}
\frac{d\sigma^{\rm frag}_{B_c^*}}{d M_{B_c+\bar c}}.
\end{equation}

It can be easy to obtain from~\cite{Bc_fragmentation_S_wave_Braaten} that:
\begin{multline}
d_{\bar b\to B_c}(M_{B_c+\bar c})=\frac{8\alpha_s |R(0)|^2M_{B_c+\bar c}}{27 \pi m_c^3} 
\int d z \;
\Theta\! \left(M_{B_c+\bar c}^2 -\frac{M_{B_c}^2}{z}-\frac{m_c^2}{1-z} \right)\\
\left( 
\frac{(1-z)(1+rz)^2rM_{B_c}^2}{(1-(1-r)z)^2(M_{B_c+\bar c}^2-m_b^2)^2}
 -\frac{(2(1-2r)-(3-4r+4r^2)z+(1-r)(1-2r)z^2)rM_{B_c}^4}
{(1-(1-r)z)(M_{B_c+\bar c}^2-m_b^2)^3} \right. \\ 
\left. -
\frac{4r^2(1-r)M_{B_c}^6}{(M_{B_c+\bar c}^2-m_b^2)^4} \right),
\end{multline}
\begin{multline}
d_{\bar b\to B_c^*}(M_{B_c+\bar c})=\frac{8\alpha_s |R(0)|^2M_{B_c+\bar c}}{27 \pi m_c^3} 
\int d z \;
\Theta \left(M_{B_c+\bar c}^2 -\frac{M_{B_c}^2}{z}-\frac{m_c^2}{1-z} \right)\\
\left(
\frac{(1-z)(1+2rz+(2+r^2)z^2)rM_{B_c}^2}{(1-(1-r)z)^2(M_{B_c+\bar c}^2-m_b^2)^2}
-\frac{(2(1+2r)-(1+12r-4r^2)z-(1-r)(1+2r)z^2)rM_{B_c}^4}{(1-(1-r)z)(M_{B_c+\bar c}^2-m_b^2)^3}
\right.  \\
\left.
-\frac{12r^2(1-r)M_{B_c}^6}{(M_{B_c+\bar c}^2-m_b^2)^4}
\right),
\end{multline}
where $M$ is  $B_c$ meson mass,  $M_{B_c+\bar c}$ is the invariant mass of $B_c$ ($B_c^*$) and $\bar c$-quark.

 Integrating over z leads to the following expressions:
\begin{multline}
d_{\bar b\to B_c}(M_{B_c+\bar c})=
\frac{4\alpha_s |R(0)|^2M_{B_c}^2r(1+r)^2 M_{B_c+\bar c}}{27 \pi m_c^3 (1-r)^4 A^2}
\times 
\\
\Biggl( 
2\left[1+2r^2+\frac{4M_{B_c}^2r(1+r)(-1+r)^2}{A}\right]
        \cdot\log\left\{\frac{A+B}{2M_{B_c+\bar c}^2}\right\}+ \\
         \frac{(1+r)^2}{A^2M_{B_c+\bar c}^4}
\biggl[
         -\bigl\{
          M_{B_c}^6(-1+r)^8(1+r)
         -M_{B_c}^4(-1+r)^4(4-14r+7r^2+r^3)M_{B_c+\bar c}^2\\
         -M_{B_c}^2(-1+r)^3(1-8r+r^2)M_{B_c+\bar c}^4
         +(2+4r-r^2+r^3)M_{B_c+\bar c}^6\bigr\}\cdot B
\\
         -2M_{B_c+\bar c}^4\bigl\{M_{B_c}^4(-1+r)^4(1-4r+2r^2)\\
         -2M_{B_c}^2(-1+r)^2(1-2r+2r^2)M_{B_c+\bar c}^2
         +(1+2r^2)M_{B_c+\bar c}^4\bigr\}
         \cdot\log\left\{\frac{A-B}{2M_{B_c+\bar c}^2}\right\}\biggr]
\Biggr)
\end{multline}

\begin{multline}
d_{\bar b\to B_c^*}(M_{B_c+\bar c})=
\frac{4\alpha_s |R(0)|^2r(1+r)^2 M_{B_c}^2M_{B_c+\bar c}}{27 \pi m_c^3(1-r)^4 A^2}\times\\
\Biggl( 
       -2\frac{1+r}{A}\bigl[M_{B_c}^2(-1+r)^2(3-8r+2r^2)-(3+4r+2r^2)M_{B_c+\bar c}^2\bigr]
      \cdot\log\left\{\frac{A+B}{2M_{B_c+\bar c}^2}\right\}
      + \\
\frac{(1+r)^2}{M_{B_c+\bar c}^4A^2}\biggl[ 
     -\bigl\{M_{B_c}^6(-1+r)^8(1+r)
      -M_{B_c}^4(-1+r)^4(2-9r^2+r^3)M_{B_c+\bar c}^2\\
      -M_{B_c}^2(-1+r)^3(-11+8r+r^2)M_{B_c+\bar c}^4
      +(12+6r-r^2+r^3)M_{B_c+\bar c}^6\bigr\}\cdot B\\
      -2M_{B_c+\bar c}^4 \bigl\{M_{B_c}^4(-1+r)^4(3-8r+2r^2)\\
      -2M_{B_c}^2(-1+r)^2(3-2r+2r^2)M_{B_c+\bar c}^2
      +(3+4r+2r^2)M_{B_c+\bar c}^4\bigr\}
      \cdot\log\left\{\frac{A-B}{2M_{B_c+\bar c}^2}\right\}\biggr]
\Biggr),
\end{multline}

where
\begin{equation}
A=(1+r)\Bigl(M_{B_c+\bar c}^2-(1-r)^2M_{B_c}^2\Bigr),
\end{equation}
and
\begin{equation}
B=(1-r)\sqrt{\Bigl(M_{B_c+\bar c}^2-(1-r)^2M_{B_c}^2\Bigr)
\Bigl(M_{B_c+\bar c}^2-(1+r)^2M_{B_c}^2\Bigr)}.
\label{eq:B}
\end{equation}
The variation of $d_{\bar b\to B_c}$ and $d_{\bar b\to B_c^*}$ with $M_{B_c+\bar c}$ are
shown in Fig.~\ref{fig:frag_minv}.

Here we face the problem how to transform the invariant mass of $B_c$ ($B_c^*$) and $\bar c$-quark 
$M_{B_c+\bar c}$ to invariant mass of $B_c$ ($B_c^*$) and $\bar D$-meson $M_{B_c+\bar D}$. Within fragmentation mechanism it naturally to assume that the $\bar D$ meson takes away the total momentum of $c$-quark, because the production of $\bar c$ quark is the last step of emission process. Such an assumption is not obvious for 
the recombination contribution. This is why two hadronization models of $\bar c$ quark have been chosen:
\begin{enumerate}
\item  $\bar D$ meson takes away the total momentum of $\bar c$-quark: $D_{c\to D}= \delta(1-z)$;
\item $\bar D$ meson takes away part of $\bar c$-quark momentum according to Kartvelishvily-Petrov-Likhoded
fragmentation function: $D_{c\to D}= z^{2.2}(1-z)$~\cite{fragmentation_KLP}.
\end{enumerate}

\begin{figure}[!t]
\centering
\resizebox*{1.\textwidth}{!}{
\includegraphics{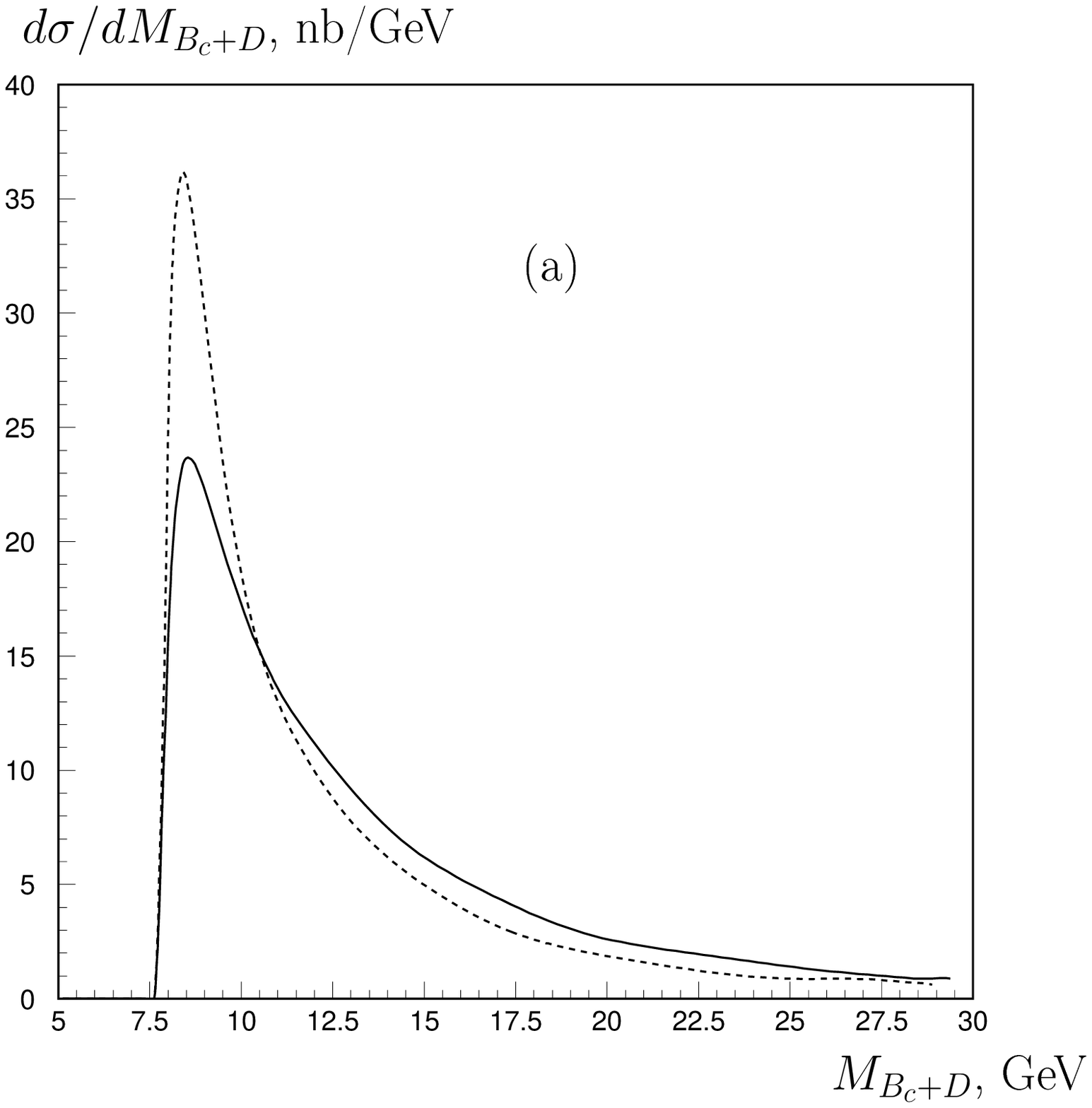}
\hfill
\includegraphics{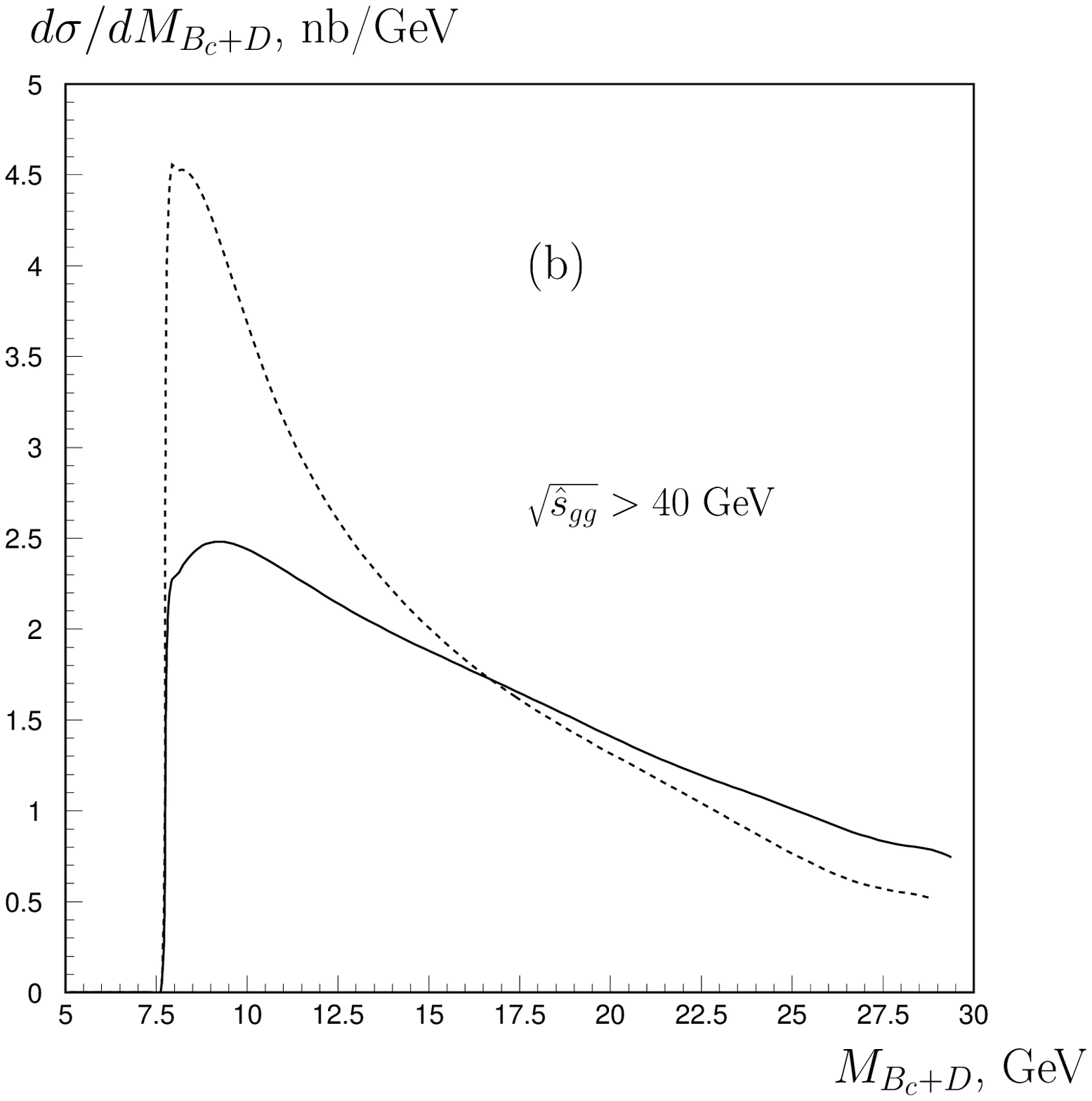}
}
\\
\parbox[t]{1.\textwidth}{
\caption{The cross section distributions over the invariant mass of $B_c$ and $\bar D$ mesons calculated within pQCD  for the process $pp \to B_c +X$ at $\sqrt{s}=14$~TeV: without cuts (a) and for $\sqrt{\hat s}>40$~GeV (b). Solid curves: $D_{c\to D}= \delta(1-z)$; dashed curves: $D_{c\to D}= z^{2.2}(1-z)$.
\hfill}
\label{fig:BcDminv_pp}}
\end{figure}
The cross section distributions over $M_{B_c+\bar D}$  are shown in Fig.~\ref{fig:BcDminv_pp} for the different kinematical regions. One can see that for the total phase space the cross section
distribution looks like the obtained within the fragmentation approach.  Nevertheless it  become essentially wider for large interaction energies, whereas within the  fragmentation approach  it  should remain the same. Thus one can conclude that the recombination contribution is dominant.

\section{The angle correlations and the decay length ratio in the associative production of $B_c$ and $D$ mesons.}

\begin{figure}[!t]
\centering
\resizebox*{1.0\textwidth}{!}{
\includegraphics{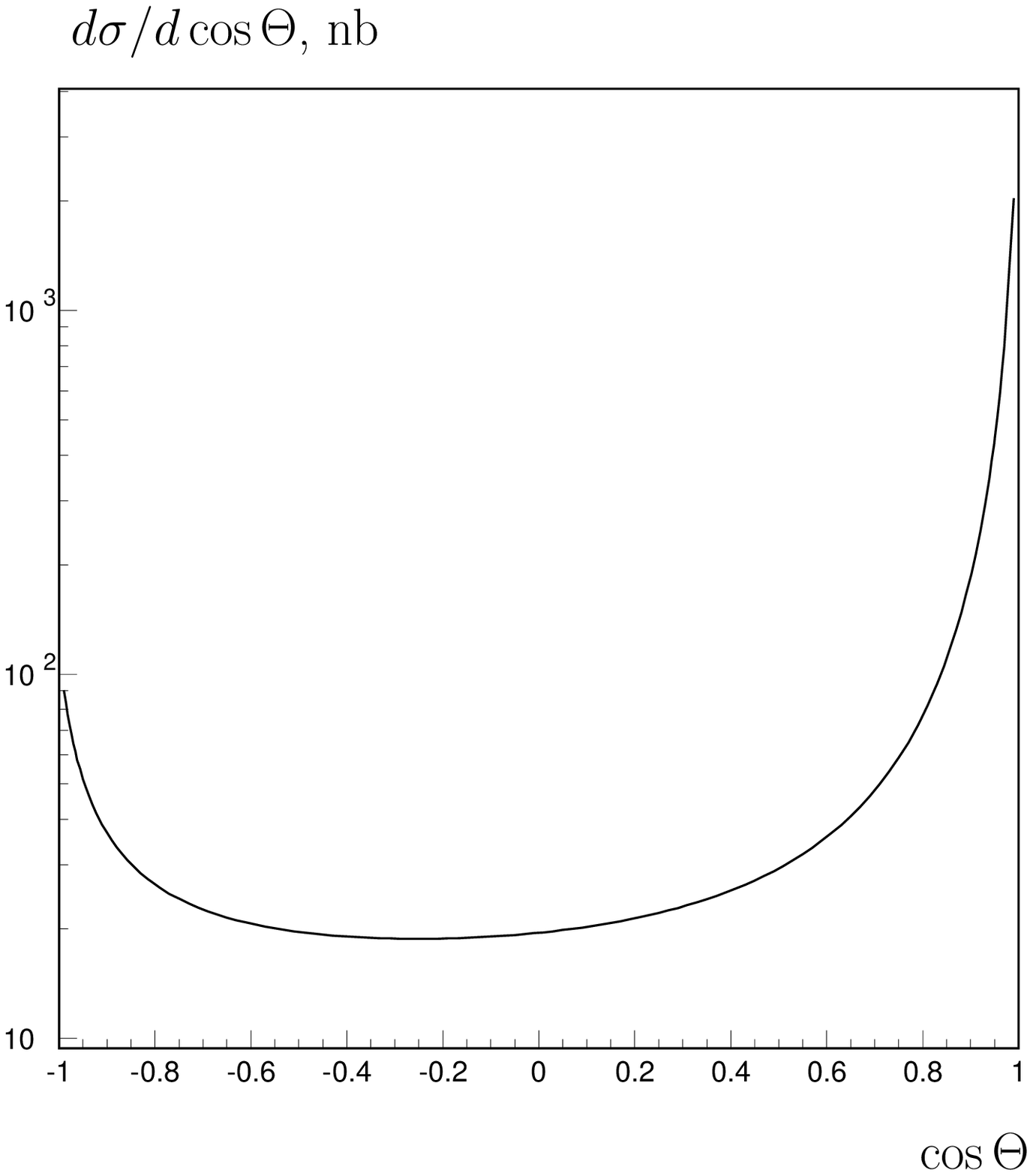}
\hfill
\includegraphics{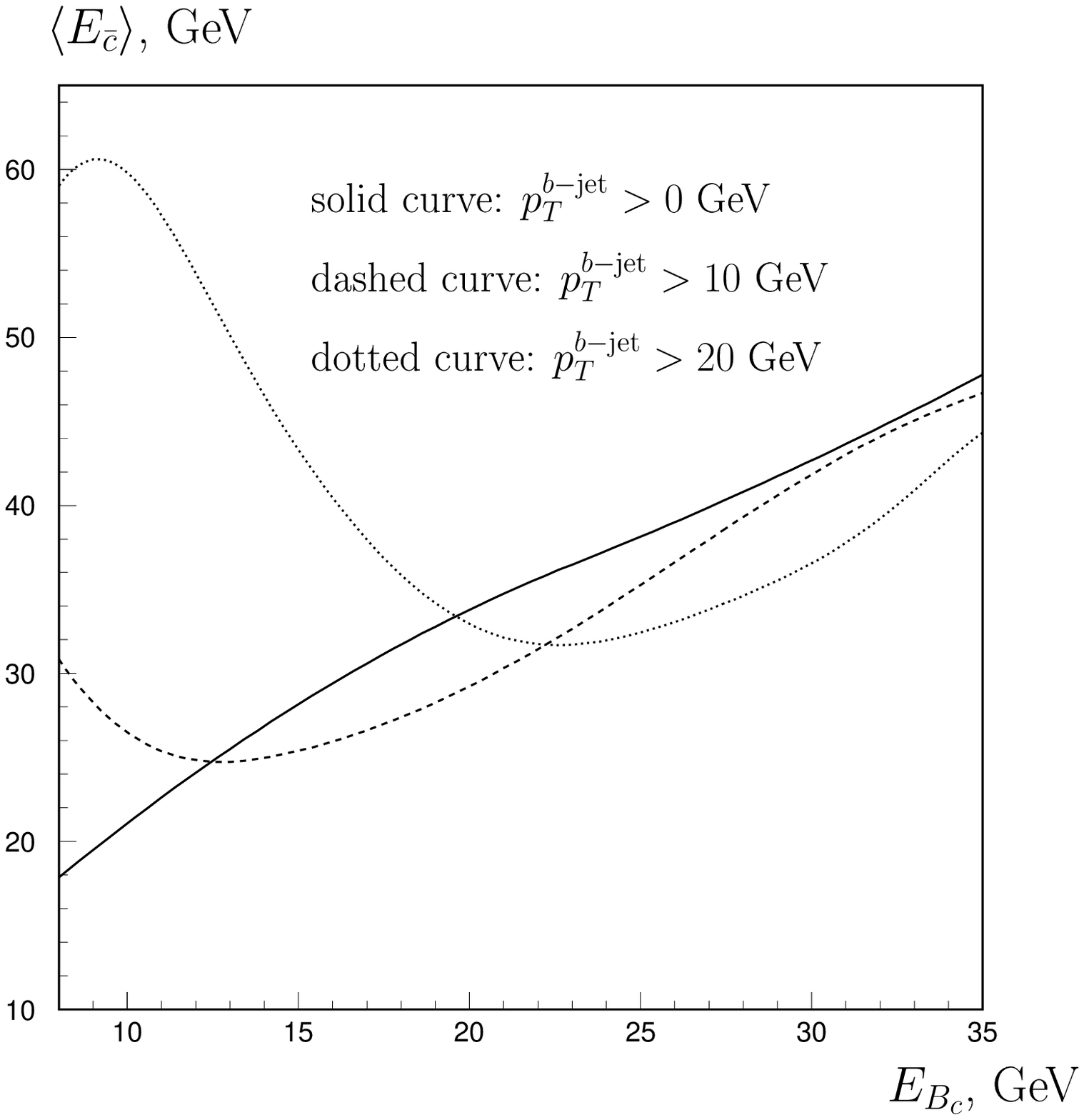}
}
\\
\parbox[t]{0.47\textwidth}{
\caption{The cross section distribution over the cosine  of angle between the  directions of motion of $B_c$ and $\bar D$ mesons predicted within pQCD  for the process $pp \to B_c +X$ at $\sqrt{s}=14$~TeV.
\hfill}
\label{fig:BcDcostheta_pp}}
\hfill
\parbox[t]{0.47\textwidth}{
\caption{The dependencies of averaged $\bar c$ quark energy on $B_c$ meson energy are represented for the different cuts on $b$-jet transverse momenta.}\label{fig:Bc_Ec_pp}}
\end{figure}

As it is shown in Fig.~\ref{fig:BcDcostheta_pp},  the cross section distribution on cosine of  the angle between the $B_c$ meson and $\bar D$ meson has the sharp maximum at
$\cos\theta\sim 1$. Moreover, about a half of $B_c$ mesons is associated by  the $\bar D$ meson moving in the close direction: $\theta \lesssim 26^o$.  Therefore, $\bar D$ meson can be used to detect $B_c$ meson.

In the Fig.~\ref{fig:Bc_Ec_pp} the dependencies of averaged $\bar c$ quark energy on $B_c$ meson energy are represented for the different cuts on $b$-jet transverse momenta. 
It can be concluded from these plots that at any $b$-jet transverse
momentum 
\begin{equation}
\langle E_{\bar c} \rangle \gtrsim 1.2 E_{B_c}. 
\end{equation}

 $\bar D$ meson we obtain that 
\begin{equation}
\langle E_D \rangle \gtrsim 1.2 E_{B_c} 
\end{equation}
for $D_{c\to D}= \delta(1-z)$
or
\begin{equation}
\langle E_D \rangle \gtrsim 0.7 E_{B_c} 
\end{equation}
 for $D_{c\to D}= z^{2.2}(1-z)$.

The decay lengths depend on particle energies and lifetimes as follows:
\begin{equation}
\langle l_D \rangle \simeq \frac{\langle E_D \rangle}{m_D} c \tau_D \qquad
 l_{B_c} \simeq \frac{ E_{B_c} }{m_{B_c}} c \tau_D.
\end{equation}

Taking into account that $\tau_D/\tau_{B_c}\simeq 2$ we obtain:
\begin{equation}
\frac{\langle l_D \rangle}{ l_{B_c} } \gtrsim 5.
\label{eq:length_ratio}
\end{equation}
This value should be compared with the fragmentation model prediction:
\begin{equation}
\frac{\langle l_D^{\rm frag} \rangle}{l_{B_c}^{\rm frag} } \sim 1 \div 2.
\end{equation}

It worth to note that the ratio $ {\langle l_D\rangle}/ l_{B_c}$ can not be used to detect the $B_c$ meson because  the distributions on $l_D$ at fixed $l_{B_c}$ very wide. Nevertheless 
the experimental measuring of this ratio could indicate the production mechanism.
 
\section{Conclusions}

The following conclusion can be drawn from the performed calculations:

\begin{enumerate}
\item
The cross section distribution over the  invariant mass of $B_c$ and $\bar D$ meson   depends essentially on kinematical cuts and can be used to research $B_c$ production mechanism at LHC. 
\item 
In many cases the $B_c$ and $\bar D$ mesons move in close directions. It could be useful to detect $B_c$ meson.
\item
The energies of $B_c$ and $\bar D$ mesons are comparable. The decay length of $\bar D$ meson by more than $5$ times larger than the decay length of $B_c$ meson. The experimental research of the ratio  between $B_c$ meson and $\bar D$ meson energies could shed light on  $B_c$ production mechanism.
\end{enumerate}

Authors thank E.~E.~Boos, L.~K.~Gladilin, N.~V.~Nikitin  and S.~P.~Baranov for the fruitful discussion.
The work of A. Berezhnoy was partially supported by Federal Agency for Science and Innovation under state contract 02.740.11.0244.

\end{document}